\documentclass{article}
\usepackage{spconf,amsmath,graphicx}

\usepackage{algorithm}
\usepackage{algorithmic}
\usepackage{bm}
\usepackage{subcaption}
\usepackage{comment}
\usepackage{xcolor}
\usepackage{cite}

\usepackage{amsmath}
\usepackage{amssymb}
\usepackage{amsfonts}

\title{Contextualized Automatic Speech Recognition \\with Attention-Based Bias Phrase Boosted Beam Search}
\name{Yui Sudo$^1$, Muhammad Shakeel$^1$, Yosuke Fukumoto$^1$, Yifan Peng$^2$, Shinji Watanabe$^2$}
\address{
  $^1$Honda Research Institute Japan Co., Ltd., Saitama, Japan\\
  $^2$Carnegie Mellon University, Pittsburgh, PA, USA}
\begin{document}
\ninept
\maketitle
\begin{abstract}
End-to-end (E2E) automatic speech recognition (ASR) methods exhibit remarkable performance. However, since the performance of such methods is intrinsically linked to the context present in the training data, E2E-ASR methods do not perform as desired for unseen user contexts (e.g., technical terms, personal names, and playlists). Thus, E2E-ASR methods must be easily contextualized by the user or developer. This paper proposes an attention-based contextual biasing method that can be customized using an editable phrase list (referred to as a bias list). The proposed method can be trained effectively by combining a bias phrase index loss and special tokens to detect the bias phrases in the input speech data. In addition, to improve the contextualization performance during inference further, we propose a bias phrase boosted (BPB) beam search algorithm based on the bias phrase index probability. Experimental results demonstrate that the proposed method consistently improves the word error rate and the character error rate of the target phrases in the bias list on both the Librispeech-960 (English) and our in-house (Japanese) dataset, respectively.
\end{abstract}
\section{Introduction}
\label{sec:intro}
% Back ground
End-to-end (E2E) automatic speech recognition (ASR)  \cite{prabhavalkar2023end,li2022recent} methods directly convert acoustic feature sequences to token sequences without requiring the multiple components used in conventional ASR systems, such as acoustic models (AM) and language models (LM). 
Various E2E-ASR methods have been proposed previously, including connectionist temporal classification (CTC) \cite{ctc1}, recurrent neural network transducer (RNN-T) \cite{rnnt1}, attention mechanism \cite{chorowski2015attention,attention1}, and their various hybrid systems \cite{watanabe2017hybrid,sainath2019two,sudo20224d}.
Since the effectiveness of E2E-ASR methods is inherently related to the context in the training data, performance expectations may not be satisfied consistently for the given user context. For example, personal names and technical terms tend to be important keywords in different contexts, but such terms may not appear frequently in the available training data, which would result in poor recognition accuracy. It is impractical to train a model for all contexts during training; thus, the user or developer should be able to contextualize the model easily without training.

A typical approach to this problem is shallow fusion using an external LM \cite{Huang2020ClassLA,47384,kojima2022,Kannan2018AnAO,sriram2018cold}. 
For example, \cite{Huang2020ClassLA,47384,kojima2022} used a weighted finite state transducer (WFST) to construct an in-class LM to facilitate contextualization for the target named entities. 
Neural LM fusion methods have been also proposed~\cite{Kannan2018AnAO,sriram2018cold}. 
The LM fusion technique attempts to enhance accuracy by combining an E2E-ASR model with an external neural LM and then rescoring the hypotheses generated by the E2E-ASR model. However, whether employing WFST or neural LMs, training an external LM requires additional training steps.

Thus, several methods have been proposed that do not require retraining. These methods include knowledge graph modeling~\cite{wang2022towards} for recognizing out-of-vocabulary named entities, contextual spelling correction~\cite{wang2022towards} using an editable phrase list, and named entity aware ASR model~\cite{sudo2023retraining} that recognize specific named entities based on phoneme similarity. 
However, these methods have limitations, such as requiring a speech synthesis (TTS) model for training and not being able to handle words other than predefined target named entities.

Deep biasing methods\cite{deepcontext2018,Jain2020ContextualRF,bruguier2019phoebe,dingliwal2023personalization} provide an alternative approach to realize effective contextualization without requiring retraining processes and TTS models. In such methods, the E2E-ASR model can be contextualized using an editable phrase list, which is referred to as a bias list in this paper. 
Most deep biasing methods implement a cross-attention layer between the bias list and input sequences to recognize the bias phrases correctly. However, it has been observed that simply adding a cross-attention layer for the bias list is not effective~\cite{huang2023contextualized}. Thus,
\cite{huang2023contextualized,han2022improving} introduced an additional branch designed to detect bias phrases, which indirectly helps to update the parameters of the cross-attention layer through an auxiliary loss.
In contrast, \cite{huber2021instant,zhou2023copyne} introduced an auxiliary loss function directly on the cross-attention layer (referred to as bias phrase index loss and will be described in Section~\ref{sec:biasdecoder}), which detects to the bias phrase index. % and allows for a direct parameter update of the cross-attention layer. 
While this approach allows for a direct parameter update of the cross-attention layer, it cannot distinguish whether the output tokens come from the bias list or not.
In addition, \cite{huber2021instant} requires two-stage training using a pretrained ASR model, which is time consuming.

This paper proposes a deep biasing method that employs both an auxiliary loss directly on the cross-attention layer, termed as bias phrase index loss, and special tokens for bias phrases to realize more effective bias phrase detection. Unlike conventional indirect methods \cite{huang2023contextualized,han2022improving}, our method facilitates the effective training of the cross-attention layer through the bias phrase index loss. Additionally, our technique departs from current methods \cite{huber2021instant} by introducing special tokens for bias phrases. This allows the model to focus on the bias phrases more effectively, eliminating the need for a two-stage training process. Furthermore, we propose a bias phrase boosted (BPB) beam search algorithm that integrates the bias phrase index probability during inference, augmenting the performance in bias phrase recognition.
The main contributions of this study are as follows:
\begin{itemize}
\vspace*{-2mm}
\leftskip -5.5mm 
\itemsep -0.5mm
    \item We propose a deep biasing model that utilizes both bias phrase index loss and special tokens for the bias phrases.
    \item We propose a bias phrase boosted (BPB) beam search algorithm to further improve the performance for the target phrases.
    \item We demonstrate that the proposed method is effective for both the Librispeech-960 and our in-house Japanese dataset.
\end{itemize}

\section{Attention-based encoder-decoder ASR}% system}
\label{sec:Preliminary}
\vspace*{-1mm}

This section describes an attention-based encoder-decoder system that consists of an audio encoder and an attention-based decoder, which are extended to the proposed method.

\subsection{Audio encoder}
\label{sec:encoder}
\vspace*{-1mm}

The audio encoder comprises two convolutional layers, a linear projection layer, and $M_{\text{a}}$ Conformer blocks \cite{gulati2020conformer}. 
%The convolutional layers subsample an audio feature sequence \begin{math}\bm{X}\end{math} into a subsampled feature sequence, and the Conformer blocks then transform the subsampled feature sequence to
The audio encoder transforms an audio feature sequence \begin{math}\bm{X}\end{math} to \begin{math}T\end{math} length hidden state vectors $\bm{H} = [\bm{h}_1, ... , \bm{h}_T] \in \mathbb{R}^{T \times d}$ where $d$ represents the dimension as follows:
\vspace*{-2mm}
\begin{align}
\label{transformer-encoder}
\bm{H} & = \mathrm{AudioEnc}(\bm{X}).
\end{align}
%Each Conformer block has two feedforward layers, a multi-headed self-attention layer, a convolution layer, and a layer-normalization layer with residual connections. 

\subsection{Attention-based decoder}
\label{sec:attention}
\vspace*{-1mm}

%In the attention-based encoder-decoder system, 
The posterior probability is formulated as follows:
\vspace*{-2mm}
\begin{equation}
\label{attlikelihood}
P_{\text{att}}(\bm{y} \mid \bm{X}) = \prod_{s=1}^{S} P\left(y_{s} \mid \bm{y}_{0:s-1}, \bm{X}\right),
\end{equation}
where $s$ and $S$ represent the token index and the total number of tokens, respectively.
Given \begin{math}\bm{H}\end{math} generated by the audio encoder in Eq. \eqref{transformer-encoder} 
and the previous token sequence \begin{math}\bm{y}_{0:s-1}\end{math}, the attention-based decoder recursively estimates the next token \begin{math}y_s\end{math} as follows:
\begin{align}
\label{eq:attndec}
P(y_s|\bm{y}_{0:s-1},\bm{X}) = \mathrm{AttnDec}(\bm{y}_{0:s-1}, \bm{H}).
\end{align}
The attention-based decoder comprises an embedding layer with a positional encoding layer, $M_{\text{d}}$ Transformer blocks, and a linear layer. Each Transformer block has a multiheaded self-attention layer, a cross-attention layer (i.e., audio attention), and a linear layer with layer normalization (LN) layers and residual connections. 
Here, the audio attention layer including the LN is formulated as follows:
\vspace*{-1mm}
\begin{equation}
\bm{U}^{'} = \mathrm{Softmax}\left(\frac{\mathrm{LN}(\bm{U}) \bm{H}^{T}}{\sqrt{d}} \right) \bm{H} + \bm{U},
\label{eq:audioattn}
\end{equation}
where $\bm{U}$ and $\bm{U}^{'}$ represent the input and output of the audio attention layer, respectively.
%Model parameters are optimized using the cross entropy loss as follows: 
%The cross entropy loss with %the one-hot vector sequence of 
%the reference transcription $\bm{y_\text{gt}}$ is used to optimize the model as follow:
%\vspace*{-1mm}
%\begin{equation}
%L_{\text{att}} =  \mathrm{CrossEntropy}(\bm{y}_{\text{gt}}, P_{\text{att}}(\bm{y} \mid \bm{X})),
%\label{eq:loss_att}
%\end{equation}
%where $\bm{y_\text{gt}}$ represents the one-hot vector sequence of the reference transcription. 
In addition, the hybrid CTC/attention model~\cite{watanabe2017hybrid} includes a CTC decoder.
%and is optimized via multitask learning with the CTC loss ($L_{\text{ctc}}$).
The attention-based decoder will be extended to the proposed bias decoder in Section~\ref{sec:biasdecoder}.

\begin{figure}[t]
\vspace*{-1mm}
    \centering
        \begin{minipage}{\textwidth}
            \includegraphics[width=0.49\textwidth]{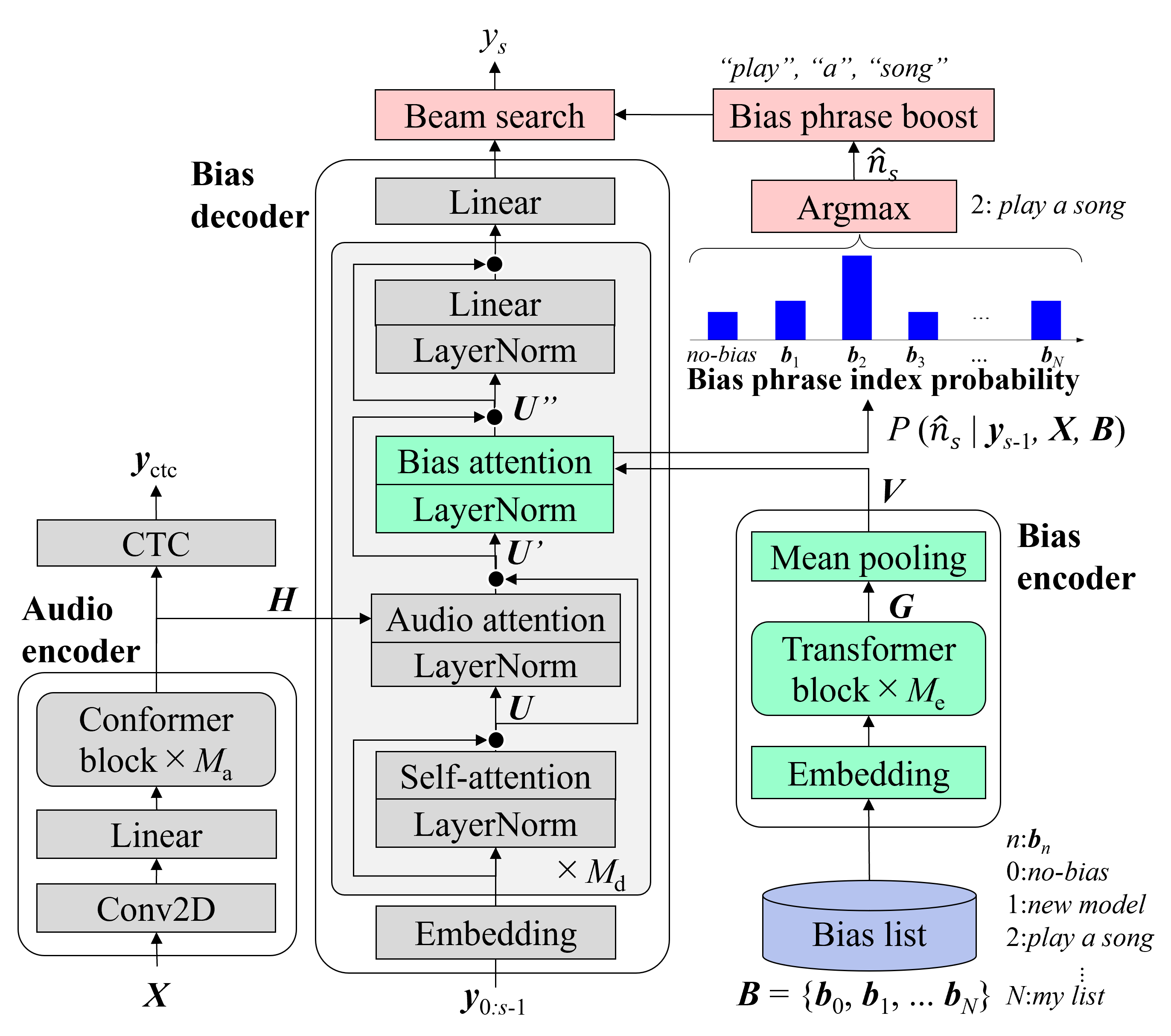} 
        \end{minipage}
        \centering
    \vspace*{-4.5mm}
    \caption{Overall architecture of the proposed method, including the audio encoder, bias encoder, and bias decoder. The BPB beam search algorithm is used during inference.}
    \label{fig:bias}
\vspace*{-4mm}
\end{figure}

\section{Proposed deep biasing method}
\label{sec:proposed}
\vspace*{-1mm}

Figure \ref{fig:bias} shows the overall architecture of the proposed method, which comprises the audio encoder, bias encoder, and bias decoder. These components are described in the following subsections.

\subsection{Bias encoder}
\label{sec:biasencoder}
\vspace*{-1mm}

The bias encoder comprises an embedding layer with a positional encoding layer, 
$M_{\text{e}}$ Transformer blocks, a mean pooling layer, and a bias list $\bm{B} = \{\bm{b}_{0}, \bm{b}_{1}, \cdots, \bm{b}_{N}$\}, where $n$ and $\bm{b}_{n}$ represent the bias phrase index and the token sequence of the $n$-th bias phrase (e.g., ``\textit{play a song}''), respectively. Here, $\bm{b}_{0}$ is a dummy phrase which means ``\textit{no-bias}''.
After applying zero padding based on the max token length $L_{\text{max}}$ in the bias list $\bm{B}$, the embedding layer and the Transformer blocks extract a set of token-level feature sequences, $\bm{G} \in \mathbb{R}^{(N+1) \times L_{\text{max}} \times d}$ as follows:
 %$ = \{\bm{G}_{0}, \bm{G}_{1}, \cdots, \bm{G}_{N}\} \in \mathbb{R}^{(N+1) \times L_{\text{max}} \times d}$ as follows:
\vspace*{-1mm}
\begin{align}
\bm{G} = \mathrm{Transformer}(\mathrm{Embedding}(\bm{B})).
\end{align}
Then, mean pooling is performed to extract a phrase-level feature sequence, $\bm{V} = [\bm{v}_{0}, \bm{v}_{1}, \cdots, \bm{v}_{N}] \in \mathbb{R}^{(N+1) \times d}$, as follows:
\vspace*{-1mm}
\begin{align}
\bm{V} = \mathrm{MeanPool}(\bm{G}).
\label{eq:hp}
\end{align}

\subsection{Bias decoder}%attention-based score loss and bias tokens}
\label{sec:biasdecoder}
\vspace*{-1mm}

The bias decoder is an extension of the attention-based decoder described in Section~\ref{sec:attention}, where an additional cross-attention layer (i.e., bias attention) is introduced to each Transformer block, as shown in Figure~\ref{fig:bias}.
Unlike  Eq.~\eqref{attlikelihood}, the posterior probability is formulated using the bias list $\bm{B}$ as follows:
\vspace*{-2mm}
\begin{equation}
\label{biaslikelihood}
P_{\text{batt}}(\bm{y} \mid \bm{X}, \bm{B})=\prod_{s=1}^{S} P\left(y_{s} \mid \bm{y}_{0:s-1}, \bm{X}, \bm{B}\right).
\end{equation}
Given $\bm{H}$, $\bm{V}$ in Eqs. \eqref{transformer-encoder}, \eqref{eq:hp}, and $\bm{y}_{0:s-1}$, the bias decoder estimates the next token \begin{math}y_s\end{math} recursively, unlike Eq.~\eqref{eq:attndec}, as follows:
\vspace*{-1mm}
\begin{align}
\label{eq:tokenprob}
P\left(y_{s} \mid \bm{y}_{0:s-1}, \bm{X}, \bm{B}\right) = \mathrm{BiasDec}(\bm{y}_{0:s-1}, \bm{H}, \bm{V}).
\end{align}
In the Transformer block of the bias decoder, the bias attention layer including the LN is formulated as follows:
\vspace*{-1mm}
\begin{equation}
\bm{U}^{''}_{} = \mathrm{Softmax}\left(\frac{\mathrm{LN}(\bm{U}^{'}) \bm{V}^{T}}{\sqrt{d}} \right) \bm{V} + \bm{U}^{'}.
\label{eq:attention}
\end{equation}
In addition, the bias attention layer estimates the bias phrase index sequence $\bm{\hat{n}} = [\hat{n}_1, \hat{n}_2, \cdots, \hat{n}_S]$ as follows:
\vspace*{-2mm}
\begin{equation}
\label{phraselikelihood}
P_{\text{bidx}}(\bm{\hat{n}} \mid \bm{X}, \bm{B})=\prod_{s=1}^{S} P\left(\hat{n}_{s} \mid \bm{y}_{0:s-1}, \bm{X}, \bm{B}\right),
\end{equation}
\vspace*{-2mm}
\begin{equation}
P\left(\hat{n}_{s} \mid \bm{y}_{0:s-1}, \bm{X}, \bm{B}\right) =  \mathrm{Softmax}\left(\frac{\mathrm{LN}(\bm{u}^{'}_{s}) \bm{V}^{T}}{\sqrt{d}} \right),
\label{eq:phraseprob}
\end{equation}
where $\bm{u}^{'}_{s}$ denotes the $s$-th feature vector of $\bm{U}^{'} = [\bm{u}^{'}_{0}, \bm{u}^{'}_{1}, \cdots, \bm{u}^{'}_{S}]$. 
%where $\bm{u}^{'}_{s}$ and $n_{s}$ represent the $s$-th feature vector of $\bm{U}^{'} = [\bm{u}^{'}_{0}, \bm{u}^{'}_{1}, \cdots, \bm{u}^{'}_{S}]$ and the bias phrase index including the no-bias option, respectively. 
For example, if a bias phrase, ``\textit{play a song}'' with a bias index of 2 (Figure~\ref{fig:bias}) is detected in a complete utterance, ``\textit{I play a song today}", the bias phrase index sequence $\bm{\hat{n}}$ = [0, 2, 2, 2, 0].
Model parameters are optimized using the cross entropy losses as follows: 
\vspace*{-1mm}
\begin{equation}
L_{\text{batt}} =  \mathrm{CrossEntropy}(\bm{y}_{\text{gt}}, P_{\text{batt}}(\bm{y} \mid \bm{X}, \bm{B})),
\label{eq:loss_bias}
\end{equation}
\vspace*{-3mm}
\begin{equation}
\label{eq:loss_score}
L_{\text{bidx}} = \mathrm{CrossEntropy}(\bm{\hat{n}}_{\text{gt}}, P_{\text{bidx}}(\bm{\hat{n}} \mid \bm{X}, \bm{B})),
\end{equation}
where $\bm{y_\text{gt}}$ and $\bm{\hat{n}}_{\text{gt}}$ represent the one-hot vector sequences of the reference transcription and the reference bias phrase index including the \textit{no-bias} option.
Here, we refer to $L_{\text{bidx}}$ as bias phrase index loss, respectively.

\subsection{Training}
\label{sec:training}
\vspace*{-1mm}

During the training process, a bias list $\bm{B}$ is created randomly from the corresponding reference transcriptions for each batch. Specifically, 0 to $N_{\text{utt}}$ bias phrases of 2 to $L_{\text{max}}$ token lengths are extracted uniformly for each utterance, resulting in a total of $N$ bias phrases ($N_{\text{utt}} \times n_{\text{batch}}$).
After the bias list $\bm{B}$ is extracted randomly, special tokens  ($<$sob$>$/$<$eob$>$) are inserted before and after the extracted phrases in the reference transcription to distinguish whether the output tokens come from the bias list or not.
The proposed method is optimized via multitask learning using the weighted sum of losses, as expressed in Eqs.~\eqref{eq:loss_bias}, \eqref{eq:loss_score}, and the CTC loss ($L_{\text{ctc}}$):
\vspace*{-1mm}
\begin{equation}
L =  \lambda_{\text{ctc}} L_{\text{ctc}} + \lambda_{\text{batt}} L_{\text{batt}} + \lambda_{\text{bidx}} L_{\text{bidx}},
\label{eq:multitask}
\end{equation}
where $\lambda_{\text{ctc}}$, $\lambda_{\text{batt}}$, and $\lambda_{\text{bidx}}$ represent the training weights.

\begin{figure}[t]
    \centering
        \begin{minipage}{0.46\textwidth}
            \includegraphics[width=\textwidth]{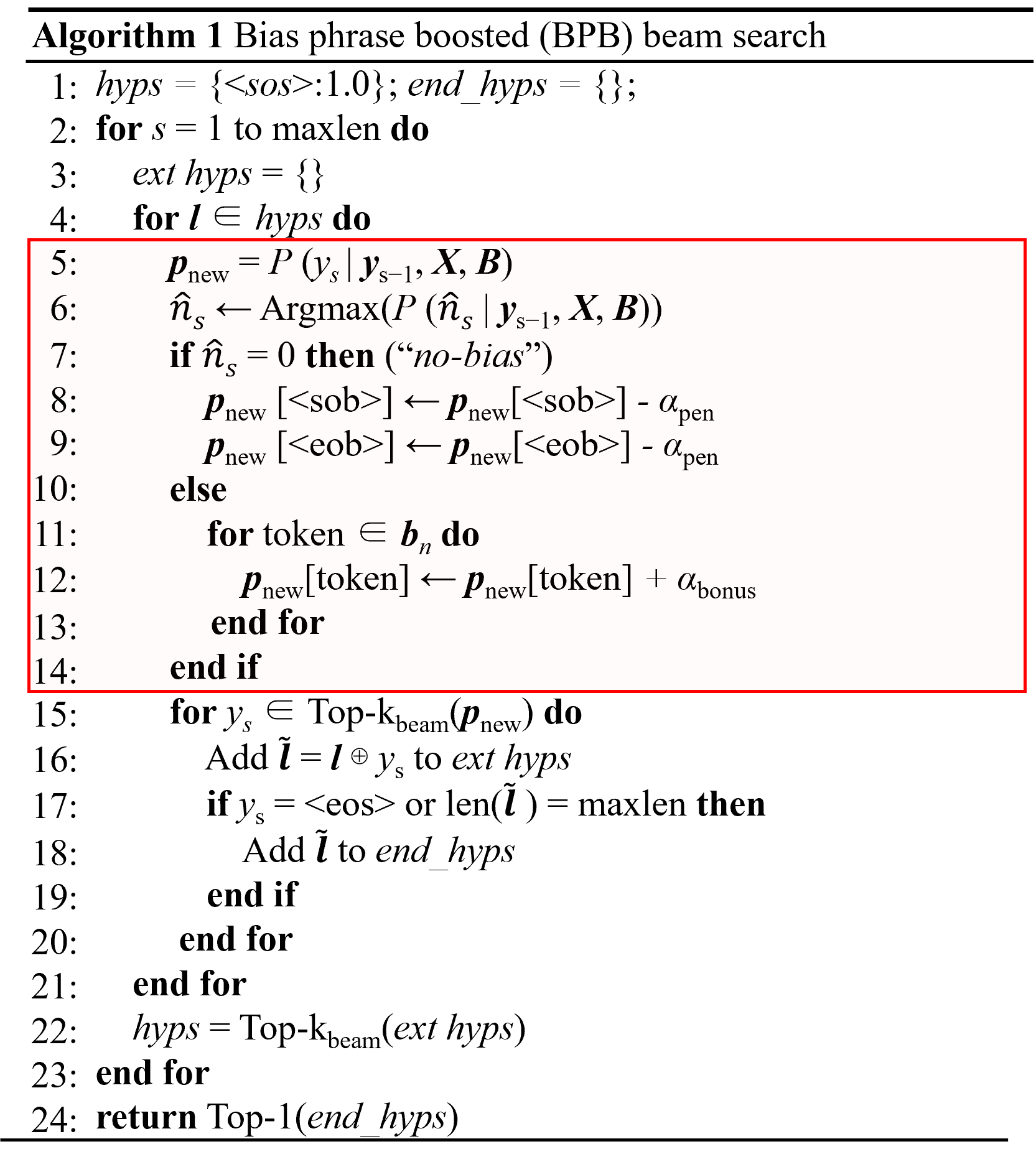} 
        \end{minipage}
        \centering
\vspace*{-5mm}
\end{figure}

\subsection{BPB beam search algorithm}
\label{sec:inference}
\vspace*{-1mm}

We also propose a bias phrase boosted (BPB) beam search algorithm that exploits the bias phrase probability as described in Algorithm 1. 
The bias decoder calculates the token probability $\bm{p}_{\text{new}}$ including the special tokens, $<$sob$>$/$<$eob$>$, using Eq.~\eqref{eq:tokenprob} (line 5).
We then estimate the bias phrase index $\hat{n}_s$ using Eq.~\eqref{eq:phraseprob} and the argmax function (line 6). 
Here, the number of bias phrases $N$ in the bias list $\bm{B}$ can increase significantly during inference, which would reduce the peak value after applying the softmax function in Eq.~\eqref{eq:attention}. Thus, Eq.~\eqref{eq:attention} is approximated using the top $k_{\text{score}}$ pruning as follows:
\vspace*{-1mm}
\begin{equation}
\bm{U}^{''} = \mathrm{Softmax}\left(\mathrm{Top\_k_{score}}\left(\frac{\mathrm{LN}(\bm{U^{'}}) \bm{V}^{T}}{\sqrt{d}}\right)\right) \bm{V} + \bm{U^{'}}.
\end{equation}
Then, if $\hat{n}_{s}$ = 0 (i.e., ``\textit{no-bias}''), the token probabilities for the special tokens $\bm{p}_{\text{new}}$[sob] and $\bm{p}_{\text{new}}$[eob] are penalized based on the weight $\alpha_\text{pen}$ (line 8, 9), otherwise, the corresponding token probabilities are increased according to the weight $\alpha_\text{bonus}$ (line 11 - 13). 
For example, if the detected bias phrase is ``\textit{play a song}'', the token probabilities for ``\textit{play}'', ``\textit{a}'', and ``\textit{song}'' are increased with $\alpha_\text{bonus}$.
Based on the boosted probabilities $\bm{p}_\text{new}$, the top $k_{\text{beam}}$ pruning is performed as in the conventional beam search~\cite{watanabe2017hybrid}.

\section{Experiment}
\vspace*{-2mm}
%Evaluation experiments were conducted to verify the effectiveness of the proposed method.

\subsection{Experimental setup}
\label{sec:experimental condition}
\vspace*{-1mm}

%The proposed method was evaluated experimentally. In this evaluation, 
The input features are 80-dimensional Mel filterbanks with a window size of 512 samples and a hop length of 160 samples. Then, SpecAugment \cite{specaug} is applied.
The audio encoder has two convolutional layers with a stride of two for downsampling, a 256-dimensional linear projection layer, and 12 Conformer blocks with 1024 linear units. 
The bias encoder and the bias decoder have three Transformer blocks with 1024 linear units and six Transformer layers with 2048 units, respectively. 
The attention layers in the audio encoder, the bias encoder, and the bias decoder are 4-multihead attentions with a dimension, $d$, of 256. 
During the training process, a bias list $\bm{B}$ is created randomly for each batch with $N_{\text{utt}}$ = 2 and $L_{\text{max}}$ = 10 described in Section~\ref{sec:training}. In this experiment, the bias list $\bm{B}$ has a total of $N$ = 50 to 200 bias phrases within a batch.
The training weights $\lambda_{\text{ctc}}$, $\lambda_{\text{batt}}$, and $\lambda_{\text{bidx}}$ (described in Eq.~\eqref{eq:multitask}) are set to 0.3, 0.7, and 1.0, respectively. The proposed model is trained for 150 epochs at a learning rate of 0.0015 with 15,000 warmup steps using the Adam optimizer.
During the decoding process, the hyper parameters of $k_{\text{beam}}$, $k_{\text{score}}$, $\alpha_\text{bonus}$, and $\alpha_\text{pen}$ (Section~\ref{sec:inference}) are set to 20, 50, 1.0 and 10.0, respectively.

The Librispeech corpus (960 h, 100 h) \cite{panayotov2015librispeech} is used to evaluate the proposed method using ESPnet as the E2E-ASR toolkit \cite{espnet}. 
The proposed method is evaluated in terms of word error rate (WER), bias phrase WER (B-WER), and unbiased phrase WER (U-WER) \cite{Le2021ContextualizedSE}.
%, which represents the biasing performance and the unbiased WER measured on words not in the bias list. 
Note that insertion errors are counted toward B-WER if the inserted phrases are present in the bias list; otherwise, insertion errors are counted toward the U-WER. 
%The bias phrases tend to be important for the end users; thus, 
The goal of the proposed method is to improve the B-WER with a slight degradation in the U-WER and overall WER.

\subsection{Preliminary analysis of the proposed techniques}
\label{sec:attn_score_loss}
\vspace*{-1mm}

\begin{table}[t]
\caption{Preliminary analysis on the Librispeech-100 test-clean.}
\vspace*{-7mm}
\label{sobeobtable}
\begin{center}
\resizebox {0.99\linewidth} {!} {
\begin{tabular}{@{}c|l|ccc}%|ccc}
\hline
ID & Model & WER & U-WER & B-WER\\
\hline
A1 & Baseline \cite{watanabe2017hybrid} & 8.59 & 5.87 & 30.71 \\
B1 & Bias decoder & 8.11 & 5.43 & 29.89 \\
B2 & B1 + bias phrase index loss & 7.53 & 5.27 & 25.92 \\ 
B3 & B2 + $<$sob$>$/$<$eob$>$ tokens & 6.93 & \textbf{4.96} & 23.00 \\
B4 & B3 + BPB beam search & \textbf{5.92} & 5.00 & \textbf{17.93} \\
\hline
\end{tabular}
}
\end{center}
\vspace*{-6mm}
\end{table}

\begin{figure}[t!]
     \centering
     \hfill
     \begin{subfigure}[b]{0.54\linewidth}
         \centering
         \includegraphics[scale=0.325]{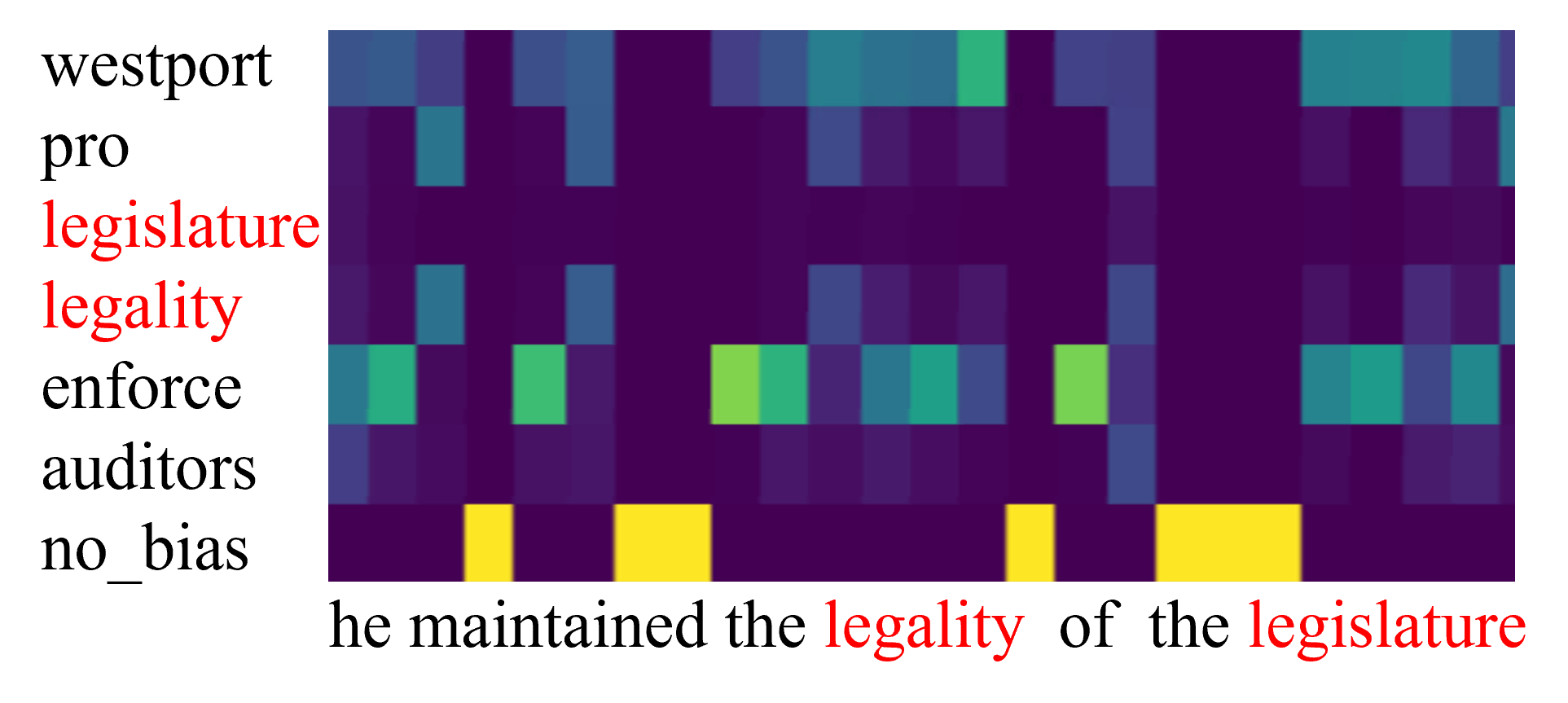}
         \vskip -0.1in
         \caption{Without bias phrase index loss}
         \label{fig:wo_attn_loss}
     \end{subfigure}
     \hfill
     \begin{subfigure}[b]{0.45\linewidth}
         \centering
         \includegraphics[scale=0.325]{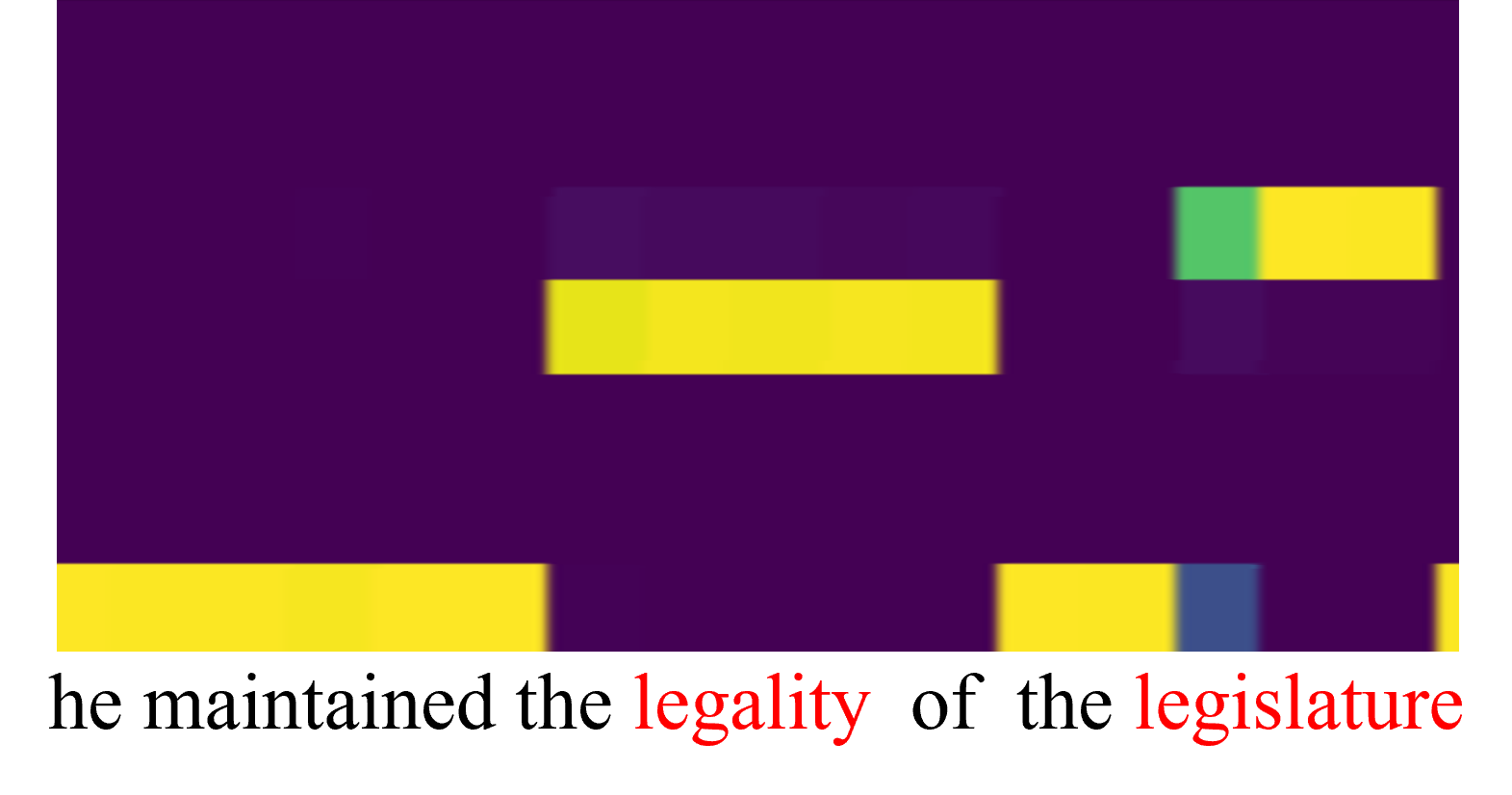}
         \vskip -0.1in
         \caption{With bias phrase index loss}
         \label{fig:w_attn_loss}
     \end{subfigure}
     \hfill
    \vskip -0.25in
    \caption{Effect of the bias phrase index loss. %, $L_{\text{bidx}}$. 
    The horizontal and vertical axes show token index $s$ and bias phrases in $\bm{B}$, respectively.}
    \label{fig:attn_score_loss}
    \vskip -0.15in
\end{figure}

\begin{table*}[t]
\caption{Main WER results obtained on Librispeech-960 data (U-WER/B-WER). \textbf{Bold} values indicate cases where the proposed method outperformed the baselines, and \underline{\textbf{underlined}} values represent the best results.}
\vspace*{-6.5mm}
\label{maintable}
\begin{center}
\resizebox {0.85\linewidth} {!} {
\begin{tabular}{@{}c|cc|cc|cc|cc}
\hline
       & \multicolumn{2}{c|}{$N$ = 0 (no-bias)} &\multicolumn{2}{c|}{$N$ = 100} & \multicolumn{2}{c|}{$N$ = 500} & \multicolumn{2}{c}{$N$ = 1000} \\%& \multicolumn{2}{c}{$N$ = 2000}  \\
Model  & test-clean & test-other & test-clean & test-other & test-clean & test-other & test-clean & test-other  \\
\hline
Baseline \cite{watanabe2017hybrid} & \textbf{3.56} & \textbf{7.55} & 3.56 & 7.55 & 3.56 & 7.55 & 3.56 & 7.55 \\
 & (\textbf{2.6}/\textbf{11.7}) & (\textbf{5.6}/\textbf{24.8}) & (2.6/11.7) & (5.6/24.8) & (\textbf{2.6}/11.7) & (5.6/24.8) & (\textbf{2.6}/11.7) & (\textbf{5.6}/24.8)\\
\hline
CPPNet \cite{huang2023contextualized} & 4.29 & 9.16  & 3.40 & 7.77 & 3.68 & 8.31 & 3.81 & 8.75 \\
 & (\textbf{2.6}/18.3) & (5.9/37.5) & (2.6/10.4) & (6.0/23.0) & (2.8/10.9) & (6.5/24.3) & (2.9/11.4) & (6.9/25.3) \\
 \hline
%Proposed & 5.81 & 9.96 & \textbf{2.86} & \underline{\textbf{5.71}} & 4.79 & \textbf{7.48} & 9.53 & 11.47 \\
%w/o one-pass  &  (4.8/13.7) & (8.0/27.1) & (\underline{\textbf{2.4}}/\textbf{6.7}) & (\underline{\textbf{4.9}}/\textbf{12.7}) & (4.1/\textbf{10.2}) & (6.2/\textbf{18.5}) & (9.0/14.1) & (10.0/\textbf{24.1}) \\
%\hline
%Proposed  & 5.05 & 8.81 & \underline{\textbf{2.81}} & \textbf{5.77} & \underline{\textbf{3.55}} & \underline{\textbf{6.40}} & 4.71 & 8.09 \\
%w/ one-pass & (3.9/14.1) & (6.6/27.9) & (\underline{\textbf{2.4}}/\underline{\textbf{6.4}}) & (\textbf{5.0}/\underline{\textbf{12.4}}) & (3.1/\underline{\textbf{7.6}}) & (\underline{\textbf{5.6}}/\underline{\textbf{13.7}}) & (4.1/\underline{\textbf{9.3}}) & (7.1/\underline{\textbf{17.2}}) \\
%\hline
Proposed & 5.81 & 9.17 & \textbf{2.94} & \textbf{6.21} & \textbf{3.24} & \textbf{6.56} & 4.07 & 7.60 \\
w/o BPB  &  (4.8/13.7) & (6.8/30.1) & (\textbf{2.5}/\textbf{6.5}) & (\textbf{5.4}/\textbf{13.1}) & (2.7/\textbf{7.9}) & (\underline{\textbf{5.5}}/\textbf{15.9}) & (3.4/\textbf{9.7}) & (6.4/\textbf{18.6}) \\
\hline
Proposed  & 5.05 & 8.81 & \underline{\textbf{2.75}} & \underline{\textbf{5.60}} & \underline{\textbf{3.21}} & \underline{\textbf{6.28}} & \underline{\textbf{3.47}} & \underline{\textbf{7.34}} \\
w/ BPB & (3.9/14.1) & (6.6/27.9) & (\underline{\textbf{2.3}}/\underline{\textbf{6.0}}) & (\underline{\textbf{4.9}}/\underline{\textbf{12.0}}) & (2.7/\underline{\textbf{7.0}}) & (\underline{\textbf{5.5}}/\underline{\textbf{13.5}}) & (3.0/\underline{\textbf{7.7}}) & (6.4/\underline{\textbf{15.8}}) \\
\hline
\end{tabular}
}
\end{center}
\vspace*{-7.5mm}
\end{table*}

Firstly, we verify the effect of the proposed techniques on the Librispeech-100 as a preliminary experiment.
Table \ref{sobeobtable} shows the effect of the bias phrase index loss, $L_{\text{bidx}}$ described in Eq.~\eqref{eq:loss_score}, the special tokens for the bias phrases ($<$sob$>$/$<$eob$>$), and the BPB beam search on the Librispeech-100 test-clean evaluation set with a bias list size of $N$ = 100.
Comparing with the baseline (the hybrid CTC/attention model~\cite{watanabe2017hybrid}), simply introducing the bias attention layer does not improve the performance (A1 vs. B1), whereas the bias phrase index loss improves the B-WER significantly, which results in an improvement to the overall WER (B1 vs. B2).
Figure \ref{fig:attn_score_loss} shows the visualization results of the bias phrase index probabilities described in Eq.~\eqref{eq:phraseprob}. The bias phrase index probabilities are estimated correctly by introducing the bias phrase index loss, $L_{\text{bidx}}$ in Eq.~\eqref{eq:loss_score}.
%This tendency was the same as that of the time-synchronous deep biasing methods reported in \cite{huang2023contextualized}.
In addition, introducing the special tokens ($<$sob$>$/$<$eob$>$) further improves the B-WER (B2 vs. B3). 
Furthermore, the BPB beam search technique significantly improves the B-WER with a slight degradation in U-WER (B3 vs. B4). 
%We consider that the phrases tend to be important for end users; thus, improving the B-WER is an extremely useful practical benefit.
%Overall, in terms of the overall WER, the results show that the proposed one-pass beam search technique obtained the best performance. 

\subsection{Main results}
\vspace*{-1mm}

Table \ref{maintable} shows the results obtained by the proposed method on the Librispeech-960 data for different bias list sizes $N$. Baseline is the hybrid CTC/attention model~\cite{watanabe2017hybrid}. 
When the bias list size $N$ = 100, the proposed method improves the B-WER, which in turn significantly improves the U-WER and WER. In addition, the proposed BPB beam search technique further improves the B-WER without degrading the overall WER and U-WER.
The B-WER and U-WER tend to deteriorate as the number of bias phrases $N$ increased; however, the proposed BPB beam search technique is particularly effective in terms of suppressing the deterioration of the B-WER.
As a result, the proposed method outperforms the baseline in terms of both WER and B-WER. 
Although the proposed method underperforms the baseline when no bias phrases are used ($N$ = 0), we do not consider it as a critical issue because the users usually register important keywords for them.

\subsection{Analysis of the BPB beam search algorithm}
\label{sec:one-pass}

\begin{figure}[t]
    \centering
        \begin{minipage}{0.34\textwidth}
            \includegraphics[width=\textwidth]{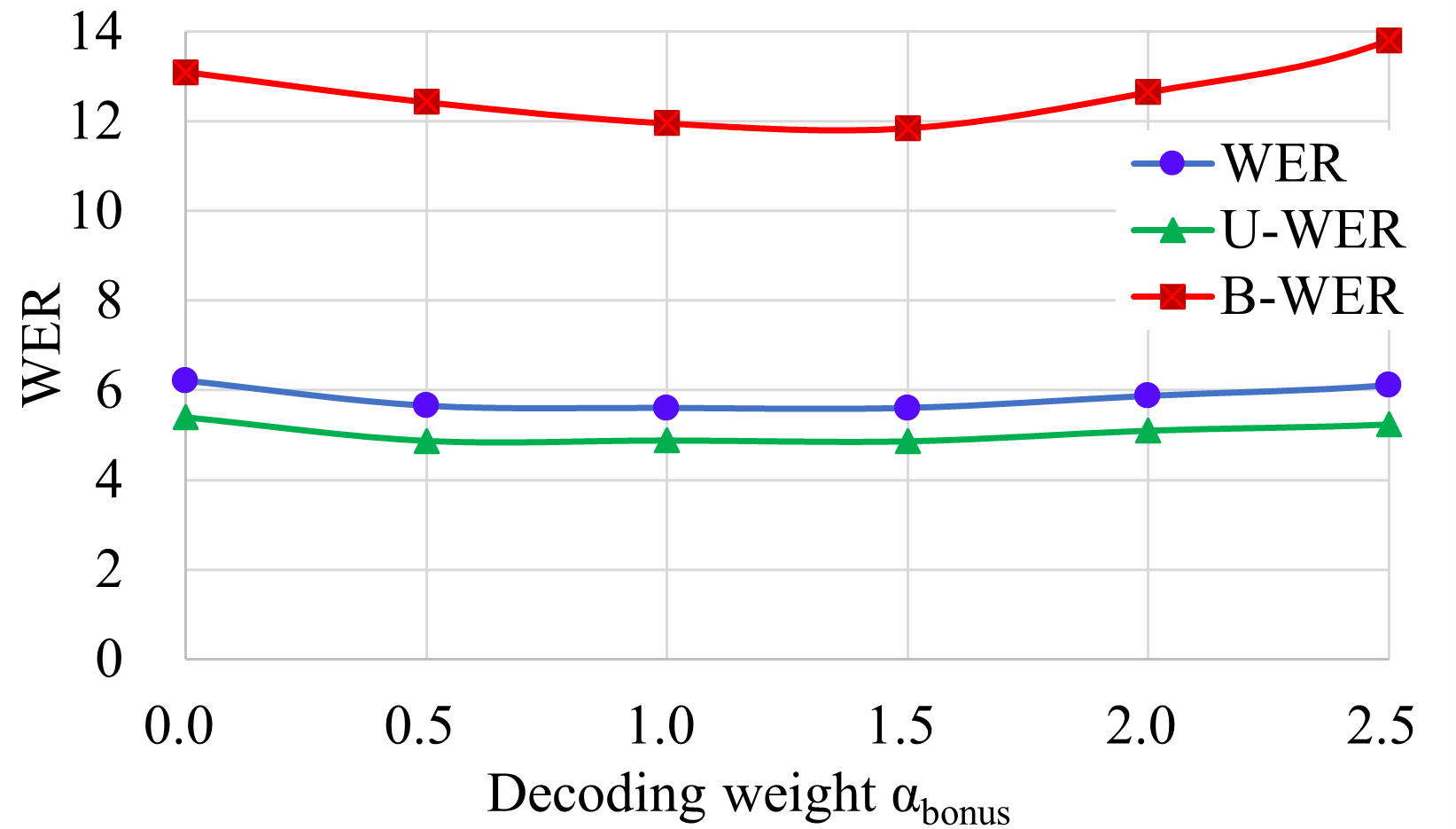} 
        \end{minipage}
        \centering
    \vspace*{-3mm}
    \caption{Effect of the decoding weight $\alpha_{\text{bonus}}$ of the BPB beam search on Librispeech-960.}
    \label{fig:onepass}
\vspace*{-3mm}
\end{figure}

Figure \ref{fig:onepass} shows the effect of the decoding weight $\alpha_\text{bonus}$ of the BPB beam search on the Librispeech-960 test-other with a bias list size of $N$ = 100. 
%The blue, green, and red lines show the overall WER, U-WER, and B-WER, respectively.
Although, even without using the proposed BPB beam search technique, the proposed method improves the B-WER as described in Section 4.3, the BPB beam search technique further improves the B-WER. When the decoding weight $\alpha_\text{bonus} > 1.5$, the B-WER, U-WER, and the overall WER deteriorate. The B-WER, U-WER, and the overall WER are the best at $\alpha_\text{bonus}$ = 1.0.

\begin{figure}[t]
    \centering
        \begin{minipage}{0.47\textwidth}
            \includegraphics[width=\textwidth]{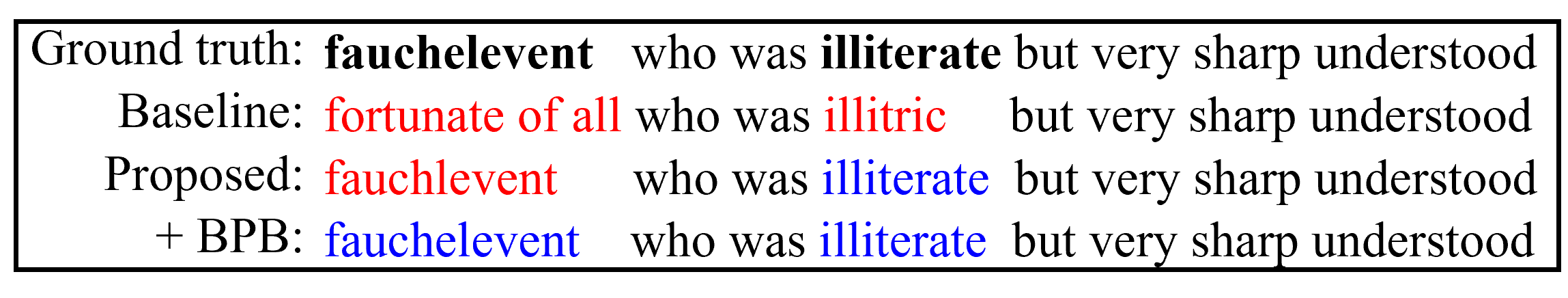} 
        \end{minipage}
        \centering
    \vspace*{-4mm}
    \caption{Typical example. Bolded faces, red and blue faces represent the bias phrases, incorrectly and correctly recognized, respectively.
    }
    \label{fig:example}
\vspace*{-0mm}
\end{figure}

Figure \ref{fig:example} illustrates the inference results from three distinct approaches: the baseline method, our proposed method excluding the BPB beam search technique, and our proposed method incorporating the BPB beam search technique.
Here, bolded face represents the bias phrases, and words in red and blue represent incorrectly and correctly recognized words, respectively.
Even without the BPB beam search technique, the proposed method reduces the misrecognition of the bias phrases compared to the baseline; however, some bias phrases are not correctly recognized even when the correct bias phrase index is estimated. In contrast, the proposed BPB beam search technique recognizes the bias phrases more correctly.

\subsection{Validation on Japanese dataset}
\label{sec:japanese}

We also validate the proposed method on our in-house dataset %\footnote{Due to privacy and confidentiality reasons, this in-house dataset will not be released publicly.} 
containing 93 hours of Japanese speech data, including meeting and morning assembly scenarios, the Corpus of Spontaneous Japanese (581 h)~\cite{csj}, and 181 hours of Japanese speech in the database developed by the Advanced Telecommunications Research Institute International~\cite{KUREMATSU1990357} with the same experimental setup described in Section~\ref{sec:experimental condition}.
Table 3 shows the evaluation results obtained on the in-house dataset when $N$ = 203 phrases, such as personal names and technical terms, are registered in the bias list $\bm{B}$. 
%Among these personal names, 45\% are not included in the training data. 
The proposed method significantly improves the B-CER with a slight degradation in the U-CER. Thus, the proposed method is effective for both English and Japanese languages.
%However, the improvement in B-CER was small compared to the results obtained on the Librispeech data because the Japanese language contains more homonyms than English; thus, unknown Kanji characters were not processed correctly when registered in the bias list. Therefore, a model that can register pronunciation information simultaneously may be developed in the future.

\begin{table}[t]
\caption{Experimental results on our in-house Japanese dataset.}
\vspace*{-6mm}
\label{castable}
\begin{center}
\begin{tabular}{@{}l|ccc}
\hline
Model & CER & U-CER & B-CER \\
%\hline
%Baseline \cite{watanabe2017hybrid} & \textbf{9.85} & \textbf{9.46} & 43.26 \\
%Proposed ($N$=0)  & 11.15 & 10.77 & 43.26 \\
%Proposed ($N$=33) & 11.07 & 10.77 & \textbf{36.17}  \\
%Proposed w/ BPB ($N$=33) & 11.08 & 10.85 & \underline{\textbf{30.50}} \\
\hline
Baseline \cite{watanabe2017hybrid} & 9.85 & \textbf{8.17} & 22.32 \\
%Proposed ($N$=0)  & 11.15 & 9.39 & 24.28 \\
Proposed ($N$=203) & \textbf{9.78} & 9.16 & \textbf{14.54} \\
Proposed w/ BPB ($N$=203) & \underline{\textbf{9.67}} & 9.20 & \underline{\textbf{13.16}}  \\
\hline
\end{tabular}
\end{center}
\vspace*{-5mm}
\end{table}

\section{Conclusion}

This study introduces a deep biasing model incorporating bias phrase index loss and specialized tokens for bias phrases. Additionally, the BPB beam search technique is employed, leveraging bias phrase index probabilities to enhance accuracy. Experimental results demonstrate that our model enhances both WER and B-WER performances. Notably, the BPB beam search boosts B-WER performance with minimal impact on overall WER, evident in both English and Japanese datasets.

\clearpage

\bibliographystyle{IEEEbib}
\bibliography{strings,references}

\end{document}